\documentclass[aoas]{imsart}

\RequirePackage{amsthm,amsmath,amsfonts,amssymb}
\RequirePackage[authoryear]{natbib}
\RequirePackage[colorlinks,citecolor=blue,urlcolor=blue]{hyperref}
\RequirePackage{graphicx}

\usepackage{bbm, bm, eqnarray}

\startlocaldefs

\usepackage{macros}

\endlocaldefs

\begin{document}

\begin{frontmatter}
\title{Modeling trajectories using functional linear differential equations}
\runtitle{Modeling trajectories using functional linear differential equations}

\begin{aug}
\author[A]{\fnms{Julia}~\snm{Wrobel}\ead[label=e1]{julia.wrobel@emory.edu}},
\author[B]{\fnms{Britton}~\snm{Sauerbrei}\ead[label=e2]{bxs561@case.edu}},
\author[B]{\fnms{Eric A.}~\snm{Kirk}\ead[label=e3]{eak152@case.edu}},
\author[C]{\fnms{Jian-Zhong}~\snm{Guo}\ead[label=e4]{jayzhong@email.unc.edu}},
\author[C]{\fnms{Adam}~\snm{Hantman}\ead[label=e5]{adam\_hantman@med.unc.edu}}
\and
\author[D]{\fnms{Jeff}~\snm{Goldsmith}\ead[label=e6]{jeff.goldsmith@columbia.edu}}
\address[A]{Department of Biostatistics and Bioinformatics,
Emory University\printead[presep={,\ }]{e1}}

\address[B]{Department of Neurosciences,
Case Western Reserve University\printead[presep={,\ }]{e2,e3}}

\address[C]{Department of Cell Biology and Physiology,
UNC Medical Center\printead[presep={,\ }]{e4,e5}}

\address[D]{Department of Biostatistics,
Columbia University Mailman School of Public Health\printead[presep={,\ }]{e6}}
\end{aug}

\begin{abstract}
We are motivated by a study that seeks to better understand the dynamic relationship between muscle activation and paw position during locomotion. For each gait cycle in this experiment, activation in the biceps and triceps is measured continuously and in parallel with paw position as a mouse trotted on a treadmill. We propose an innovative general regression method that draws from both ordinary differential equations and functional data analysis to model the relationship between these functional inputs and responses as a dynamical system that evolves over time. Specifically, our model addresses gaps in both literatures and borrows strength across curves estimating ODE parameters across all curves simultaneously rather than separately modeling each functional observation. Our approach compares favorably to related functional data methods in simulations and in cross-validated predictive accuracy of paw position in the gait data. In the analysis of the gait cycles, we find that paw speed and position are dynamically influenced by inputs from the biceps and triceps muscles, and that the effect of muscle activation persists beyond the activation itself.
\end{abstract}

\begin{keyword}
\kwd{functional regression}
\kwd{ordinary differential  equations}
\kwd{nonlinear least squares}
\kwd{dynamical systems}
\end{keyword}

\end{frontmatter}



\section{Introduction}
\label{sec:intro}

Our motivating data come from a study that collected upward trajectories of paw position during gait cycles as a mouse trotted on a treadmill; the paw positions during each gait cycle were measured concurrently with muscle activity in the biceps and triceps, areas of the forelimb known to be important for locomotion. These data were collected in an effort to understand the relationship between muscle activity and paw position during gait cycles. Recent work using these and similar data suggests that the dynamics of the paw during dexterous, voluntary movements are tightly coupled to neuromuscular control signals \citep{guo2015, sauerbrei2020, becker2020, kirk2023output}. This is an example from the increasingly common class of problems where outcomes and predictors are measured densely in parallel. For these data streams, the goal is to understand the relationship between inputs and outputs that are both functions measured on the same domain.

To better quantify how muscle activity affects current and future paw position during a gait cycle, we need a method that (1) allows future position to depend on past but not future muscle activation, (2) allows future position to be affected by initial position, (3) has parameters that model the relationship between the paw position and biceps and triceps muscles as a dynamical system of inputs and outputs, the state of which evolves over time, and (4) can accommodate repeated functional observations across trials. These problems cannot be simultaneously addressed by current methods, in large part because existing estimation techniques for dynamical systems can only be applied to a single realization of a trajectory rather than repeated observations. We therefore develop a novel regression framework that combines ordinary differential equations (ODEs) and functional regression and is well-suited to address the problems our data presents. This work is connected to both the ODE and functional data analysis literatures, which we review in Sections \ref{sec:odes} and \ref{sec:fda}, respectively. First, in Sections \ref{sec:data} and \ref{sec:flode}, we describe our motivating data and model structure in more detail.


\subsection{Gait analysis data}
\label{sec:data}

We present data that were collected as part of a study examining how neuromuscular activation is involved in enacting coordinated movement. Several related experiments, including \cite{guo2015}, \cite{pistohl2008prediction}, and \cite{sauerbrei2020}, have shown that the motor cortex generates a continuous signal driving skilled movements; our study examines how muscle activity influences limb kinematics during locomotion.

In the experiment that generated our data, a single mouse was trained to trot on a treadmill at approximately 20 cm/s. Video recordings of the task completion were used to extract upward trajectories of paw position of the front left paw for each gait cycle. Intramuscular electromyograms (EMG) were recorded from forelimb flexor (biceps brachii) and extensor (triceps brachii) muscles and processed using standard techniques.
The primary objectives of this study were to investigate how muscle activity during locomotion adapts to changes in mechanical load and to explore how these adjustments are represented in and influenced by brain activity patterns. The study concluded that while the motor cortex, a brain region involved in voluntary movement, retains a representation of limb loads, it is not essential for adjusting muscle activity to counteract these loads. This conclusion is drawn from invasive brain recordings that are not feasible in humans, although similar findings have been observed in nonhuman primates performing voluntary upper limb movements. More details on data collection are provided in \cite{kirk2023output}.

For our analysis, each trial of the experiment is a single gait cycle, and our data include 161 gait cycles in total. In this work, we define a gait cycle to begin when the paw comes to rest on the treadmill, continue through the lifting motion, and end when the paw has again come to rest on the treadmill.

\begin{figure}[h]
  \centering
     \begin{tabular}{cc}
      	\includegraphics[width= .95\textwidth]{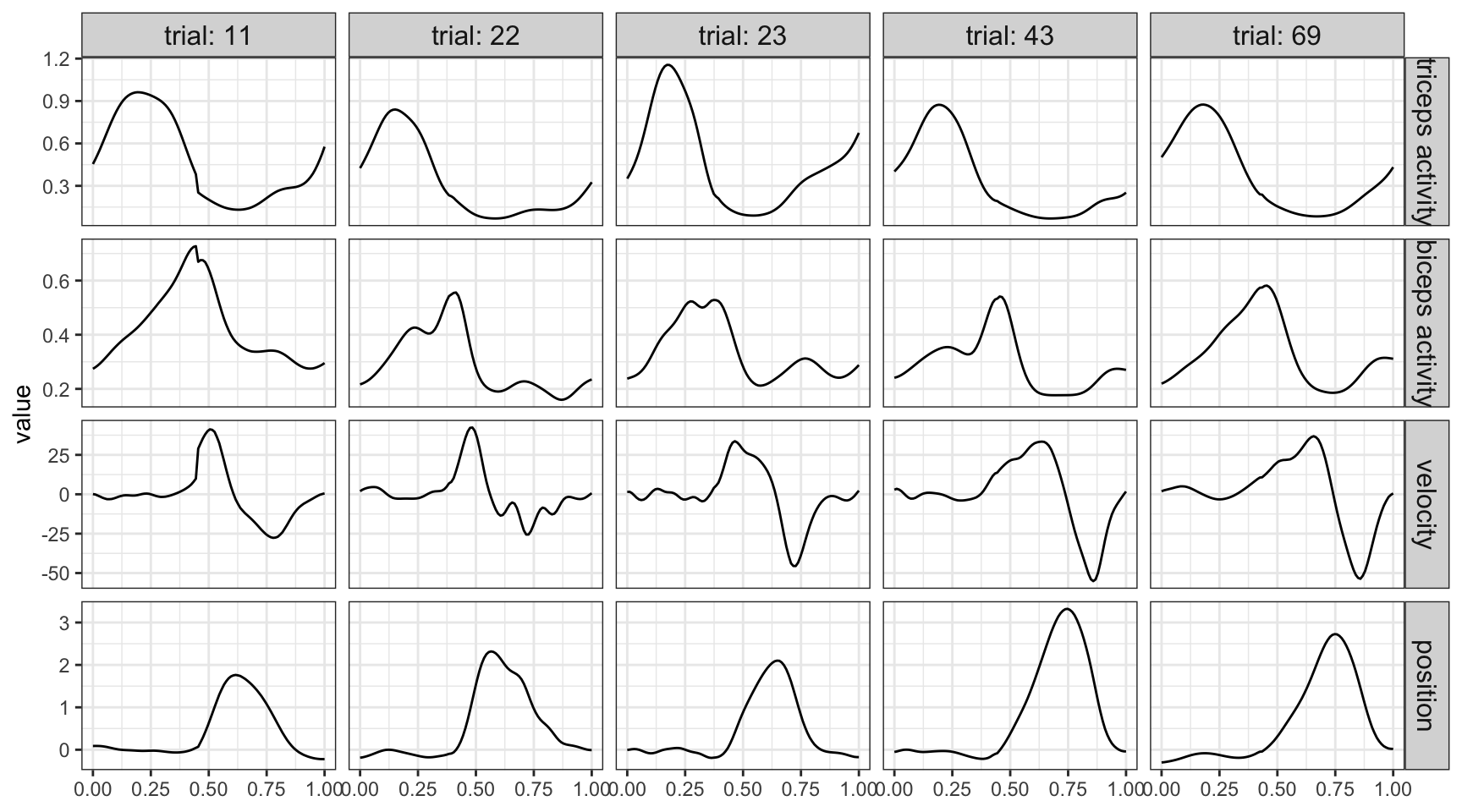}
     \end{tabular}
     \caption{Smoothed triceps activation (first row), smoothed biceps activation (second row), paw upward velocity (third row), and paw position (fourth row) for five randomly selected gait cycles.
}
    \label{fig:flode_data}
\end{figure}

Figure (\ref{fig:flode_data}) shows smoothed triceps activation (first row), smoothed biceps activation (second row), paw upward velocity (third row), and paw position (fourth row) for five randomly selected gait cycles. Taken together, Figure (\ref{fig:flode_data}) illustrates the dynamics of the muscle activation and paw position during a gait cycle. At the outset of each gait cycle, the paw is on the treadmill and the triceps are active to support the animal. In the early stages of the gait cycle (e.g. $t \in [0, .4]$), activation in the triceps decreases while the biceps muscle begins to become active; these changes induce the paw upward speed to increase, which in turn raises the paw position. In the next stage of each gait cycle  ($t \in [.5, .75]$), the biceps activation drops and, as the paw returns to rest on the treadmill at the end of the gait cycle, the tricep muscle is reactivated. In addition, we note that paw position in the early phase of each gait cycle depends on the initial paw position, although this dependence decreases as time increases.


\subsection{The flode model}
\label{sec:flode}

The biological underpinnings of our data are a dynamical system in which values of interrelated components are constantly evolving. These components include the current paw position, speed, and direction, as well as muscle activation patterns, and they come together to guide the paw velocity and position over the course of a gait cycle. More explicitly, the paw is being acted on by outside forces coming from the flexor and extensor muscles, and these forces drive changes in velocity of the paw which then influences position. Collectively, the initial paw position, the current and recent the state of the muscular system, and unmeasured additional forces combine in a way that is nonlinear and time-varying to produce the instantaneous state of the paw at each point in time. Ordinary differential equations (ODEs) can be used to model and succinctly describe dynamical systems and their evolution over time, but no framework currently exists in which ODEs can be used to model or borrow information across repeated i.i.d. trajectories.


We introduce the \textit{flode} (\textbf{f}unctional \textbf{l}inear \textbf{o}rdinary \textbf{d}ifferential \textbf{e}quation) model, a novel functional regression framework that represents this neurobiological system of inputs (muscle activity) and outputs (paw position during a gait cycle). The \textit{flode} model is both an ordinary differential equation, which reflects the dynamic nature of our data and allows us to incorporate how change in paw position influences position at time $t$, and a functional regression model, which improves estimation by borrowing strength across repeated trajectories. Like other first order ODEs \textit{flode} has both a differential and an integrated form. In its differential form, our model is

\begin{equation}
\label{eq:flode_deriv}
y'_i(t) = -\alpha y_i(t) + \delta_i(t) + \mathcal{B}_0(t) + \sum_{p = 1}^P \mathcal{B}_p(t) x_{ip}(t),
\end{equation}

\noindent
where  $y_i(t)$ and $y'_i(t)$ are the paw position and first derivative of paw position (velocity) at time $t$, $x_{ip}(t), p \in 1\ldots P$ are trial-specific muscle activation patterns, and $\alpha$, $\delta_i(t)$, and $\mathcal{B}_p(t), p\in 0\ldots P$ are parameters to be estimated from the data. The $x_{ip}(t)$ functions, analogous to covariates in a traditional regression model, can more generally be called ``forcing functions'' because they are additional input forces that act on the ODE system.

Many systems of differential equations cannot be solved analytically, which makes traditional statistical estimation techniques with the observed data $Y$ as the outcome challenging. However, the class of ODEs we consider has a solution, which we parameterize in terms of the initial value to obtain the integrated form of the \textit{flode} model

\begin{equation}
\label{eq:flode_mod}
	Y_i(t) = y_i(0)e^{-\alpha t} + \int_0^t e^{-\alpha (t-s)}\delta_i(s)ds + \sum_{p=0}^P\int_0^t e^{-\alpha (t-s)} \mathcal{B}_p(s)x_{ip}(s)ds + \epsilon_i(t).
\end{equation}

\noindent
We make a distinction between $y_i(t)$, the true (unobserved) paw position at time $t$, and $Y_i(t)$, the paw position at time $t$ observed with measurement error $\epsilon_i(t)$. Thus, in model (\ref{eq:flode_mod}) above, we assume the outcome $Y_i(t)$ is measured with error but depends on the true initial position $y_i(0)$.

The \textit{flode} model is a buffered system, meaning the response time is longer than the time interval in which the input changes. In our context, the inputs are muscle activation patterns that precede the more prolonged paw lift response. The scalar parameter $\alpha$, called the buffering parameter, indicates the amount of buffering on the system, and can be interpreted as a quantitative measure of resistance to change. As $\alpha \to 0$, buffering increases, and the effects of forcing functions and initial position persist in time. As $\alpha \in (0, \infty)$ grows larger, the effects of forcing functions and initial position becomes instantaneous. The $\mathcal{B}_p(t)$ are coefficient functions that measure the impact of changes in the forcing function $x_{ip}(t)$ on the system independent of buffering, interpreted as the change in paw velocity at time $t$, $y'_i(t)$, given a one unit change in forcing function value $x_{ip}(t)$. $\mathcal{B}_0(t)$ and $\delta_i(t)$ are the population-level and trial-specific intercepts, respectively. The random effects $\delta_i(t)$  capture residual within-trial correlation; while much of each gait cycle is known to be driven by the muscle activation in the biceps and triceps, other muscles and brain regions also contribute, and the $\delta_i(t)$ term is intended to capture changes in position driven by unmeasured forces.

\begin{figure}
\begin{center}
      	\includegraphics[width= .95\textwidth]{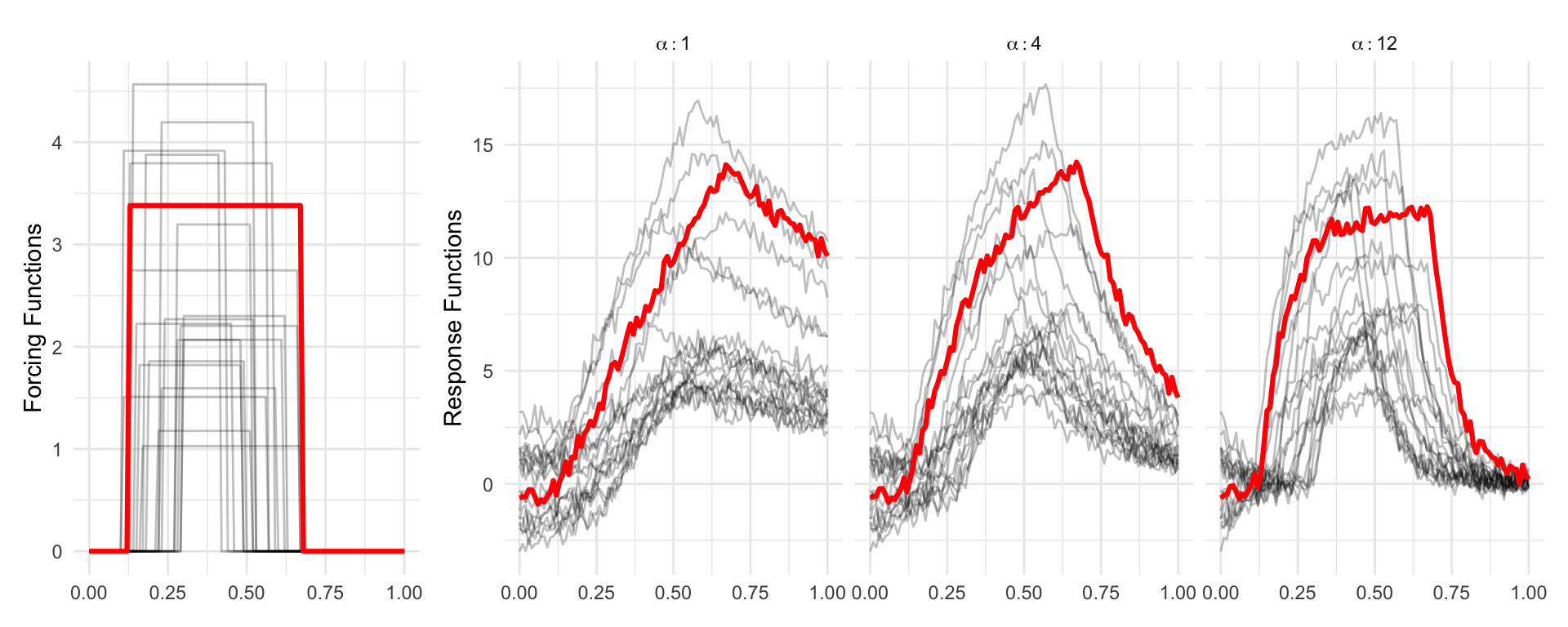}
\end{center}
     \caption{Simulated datasets with $N = 20$ trials  generated from the integrated form of the \textit{flode} model in Equation (\ref{eq:flode_mod}), with $\mathcal{B}_0(t) = 0$ and $\mathcal{B}_1(t)$ constant over $t$. The left panel shows forcing functions $x_{i}(t)$ which are shared across datasets. Remaining panels show observed position trajectories $Y_i(t)$ under $\alpha \in \{1, 4, 12\}$. The position trajectories highlighted in red correspond to the highlighted forcing function.}
    \label{fig:sim_data}
\end{figure}

Figure \ref{fig:sim_data} illustrates the \textit{flode} model by showing datasets generated from the integrated form of the \textit{flode} model in Equation (\ref{eq:flode_mod}) for three $\alpha$ values. In these simplified examples, the coefficient function $\mathcal{B}_0(t) = 0$ and $\mathcal{B}_1(t)$ is constant over $t$. Forcing functions $x_{i1}(t)$ are shared across the simulated datasets and are shown in left panel of Figure \ref{fig:sim_data}. These are generated as step functions with random step starts, ends, and heights; a single forcing function is highlighted in red. The right panels show the position trajectories $Y_i(t)$ under the integrated \textit{flode} model~(\ref{eq:flode_mod}), with the outcome corresponding to the forcing function shown in red highlighted as well. These panels build intuition for the overall behavior of the \textit{flode} model and impact of the buffering parameter $\alpha$. When $\alpha = 1$ buffering is high: the initial position is carried forward in time, and the position trajectory depends on the current and historical values the forcing function. This buffering decreases as $\alpha$ increases, and the impact of initial position and forcing functions becomes nearly instantaneous or concurrent in time. By explicitly parameterizing the degree of buffering and allowing $\mathcal{B}_0(t)$ and $\mathcal{B}_1(t)$ to be non-constant in practice, \textit{flode} provides a flexible but interpretable modeling framework.

The \textit{flode} model is a linear differential equation with several innovations to reflect the biological process that is hypothesized to generate the observed data. To familiarize readers with existing ordinary differential equations and their use in statistics, we review the ODE literature in Section \ref{sec:odes}. An overview of the functional data analysis literature with emphasis on functional-response regression models, provided in Section \ref{sec:fda}, is also pertinent since the paw trajectories are conceptualized and modeled as \textit{i.i.d.} realizations of functions that are observed over time.


\subsection{ODEs}
\label{sec:odes}

Systems of ordinary differential equations (ODEs) can be used to directly model the relationship between an outcome and its derivatives, leading to widespread popularity for modeling dynamical systems in physics, biology, neuroscience, and other disciplines. First-order ODEs, which incorporate only the first derivative of $y$, follow the general form given in Equation (\ref{eq:flode_deriv}); however, because it is uncommon to model repeated trials and account for trial-to-trial variation, first-order ODEs typically do not include the $\delta_i(t)$ terms. Equation (\ref{eq:flode_deriv}) is said to be a linear as well as first-order differential equation because its right hand side can be written as a linear combination of $y$ and terms that do not contain $y$  \citep{tennenbaum1985}. Higher order and nonlinear ODEs are possible but do not have analytic solutions, and the first order linear ODE is complex enough to capture the relationship between our outcome, its derivative, and outside forces acting on the system.
Our approach  to represent data potentially generated by higher-order models using a lower-order ODE is consistent previous literature \cite{bekins1998comparison, nelson2002mathematical, yavuzturk1999transient}.
ODEs that are analytically solvable do not have a unique solution because they contain an indefinite integral, and it is therefore common to solve in terms of a particular value. To satisfy uniqueness and obtain the integrated model in Equation (\ref{eq:flode_mod}), we solve Equation (\ref{eq:flode_deriv}) terms of the initial value $y(0)$, which is interpretable in our data setting.

Most applications of ODEs in science and engineering focus on restrictive rather than general settings. For example, a simple first-order example from chemistry is the radioactive decay model, which predicts that the quantity of a radioactive substance decreases exponentially over time at a rate proportional to its current amount \citep{loveland2017modern}.
In the past this specificity has limited their use in statistics, but they are growing in popularity. \cite{chen2017} reconstructs gene regulatory networks by estimating sparse nonlinear ODEs for noisy gene expression data, building on previous work \citep{lu2011, henderson2014}.
\cite{ramsay2017} conceptualize dynamical systems as data-driven statistical models. This book provides a framework for estimating a large class of differential equations, as well as an overview of ODE-based models that expands on earlier work from \cite{ramsay2007} for parameter estimation in nonlinear ODEs.

Recently, \cite{dai2022kernel} and \cite{dai2021post} introduced a kernel ODE model with sparsity regularization and parameter inference that is general enough to encompass more special-case ODE models previously used in the statistical literature, including \citep{lu2011, dattner2015, wu2014, chen2017}.
While this general framework is well-suited to estimate parameters for a single realization of an ODE, it does not accommodate multiple trials or the complexities that arise in such cases.


\subsection{Functional regression models}
\label{sec:fda}

Our data setting and proposed methods draw on techniques from functional data analysis. In this framework, curves $Y_i(t)$ are the fundamental unit of statistical analysis, and functional analogs of univariate methods like regression, PCA, and others build on this perspective. Functional response regression models capture the relationship between response curves $Y_i(t), i \in 1 \ldots N$ from $N$ independent trials and covariate(s) $x_i$, which can be scalar or functional. In particular, function-on-function regression allows for both functional responses and functional predictors that can be observed on different domains \citep{ramsay2005}.

Special cases of function-on-function regression in which the response and predictors are measured over the same time domain include the linear functional concurrent model \citep{fan2008, goldsmith2017} and the historical functional regression model \citep{malfait2003, yao2005, leroux2018}. The concurrent model uses the current value of the predictor to measure the response at each point in time, but doesn't allow the covariates to affect future values of the response. The functional historical model allows the response at time $t$ to be influenced only by the predictors up to time $t$; this is ideal for data where the response and predictor are measured on the same domain, and prevents future values of the predictors from influencing the present value of the response. These special cases, and in particular the historical regression, are most relevant to our motivating data and our proposed methods. Advances in functional regression and accompanying software facilitate the use of functional historical and functional concurrent regression models, allow the inclusion of functional trial-specific random effects \citep{scheipl2015, scheipl2016, refund}, and extend to nonlinear forms \citep{rao2021}.

The historical model with trial-specific random intercept $\gamma_i(t)$ is given by
\begin{equation}
\label{eq:fhist_mod}
Y_i(t) = \gamma_i(t) + \beta_0(t) +  \sum_{p = 1}^P \int_{s = 0}^t \beta_p(t,s) x_{ip}(s)ds+ \epsilon_i(t).
\end{equation}

\noindent
Here $\beta_0(t)$ is the population-level intercept, and each $\beta_p(t,s)$ is a coefficient surface that is integrated to relate the predictor to the response. This flexible model is designed to handle repeated functional observations, and inclusion of the random intercept $\gamma_i(t)$ accounts for within-trial residual correlation in the errors after modeling the relationship between the outcome and the covariates curves. Conceptually, both the integrated \textit{flode} model in (\ref{eq:flode_mod}) and functional historical regression use predictors, including their recent history, to understand current values of the response function. For each functional covariate, functional historical regression estimates the coefficient surface $\beta(t,s)$ appearing in Equation (\ref{eq:fhist_mod}). The \textit{flode} model does not estimate a coefficient surface directly, but instead estimates population-level parameters $\alpha$ and coefficient functions $\mathcal{B}_p(t)$ which control the buffering and effect of forcing functions on output velocity, respectively. As a byproduct of the \textit{flode} model specification in Equation (\ref{eq:flode_mod}), $\alpha$ and each $\mathcal{B}_p(t)$ can be combined to produce a surface of the form $e^{-\alpha (t-s)} \mathcal{B}_p(s) \mathcal{I}(s < t)$. The \textit{flode} surface $e^{-\alpha (t-s)} \mathcal{B}(s)$ has a more specific parametric structure than the historical surface $\beta(s,t)$; as a result, the historical model can overfit data that is generated by the \textit{flode} model. However, we can leverage the conceptual similarities and the resulting coefficient surfaces to compare \textit{flode} with functional historical regression in practical settings.

 The functional concurrent model with trial-specific random intercept $\gamma_i(t)$ is given by
\begin{equation}
\label{eq:fconc_mod}
Y_i(t) = \gamma_i(t) + \beta_0(t) +  \sum_{p = 1}^P \beta_p(t) x_{ip}(t)+ \epsilon_i(t).
\end{equation}

\noindent
This model is similar to the historical model in Equation \ref{eq:fhist_mod}, but less flexible in that the coefficients $\beta_p(t)$ are coefficient \textit{functions} rather than coefficient \textit{surfaces}; these $\beta_p(t)$ capture the relationship between $x_p(t)$ and $Y(t)$ at each time $t$ but do not allow the recent history of covariate functions to affect the current outcome value. Inspection of the \textit{flode} model in (\ref{eq:flode_mod}) suggests that as $\alpha$ increases, the impact of a forcing function's recent history will decrease. This observation, which is consistent with the illustration in Figure~\ref{fig:sim_data}, suggests that it can be useful to compare \textit{flode} to concurrent regression in practice.

From a conceptual standpoint, \textit{flode} introduces a new framework for thinking about the relationship between inputs and outputs in an ODE system, and no existing functional regression models offer this interpretation. Initial position is a crucial element of the \textit{flode} framework because it provides a specific analytic solution to the ODE in Equation (\ref{eq:flode_deriv}); in contrast, initial position is not a natural element of  the historical or concurrent models and does not have precedent in the functional regression literature. Explicitly incorporating initial position into a functional regression context is both critical for our dynamical systems approach and a novel contribution in its own right. Finally, the \textit{flode} model is nonlinear in its parameter $\alpha$, a development which other functional regression methods haven't directly addressed.


\section{Methods}
\label{sec:methods}

Our work introduces models (\ref{eq:flode_deriv}) and (\ref{eq:flode_mod}), a novel framework for modeling functional observations with an explicit dynamical systems interpretation.


\subsection{Data processing}

Quality control on kinematics and EMG was performed at multiple stages, as outlined in \cite{kirk2023output}. First, the kinematic output was closely examined to ensure accurate pose estimates from each camera. This process included calculating reprojection errors from the estimated 3D positions back onto the individual camera views and visually inspecting stride-aligned limb kinematics. Second, detected step cycles were excluded from analysis if the swing phase duration was less than 60 ms or more than 400 ms. Third, EMG signals from the biceps and triceps were examined offline to verify the signal-to-noise ratio and ensure the absence of significant crosstalk between electrodes and contamination from antagonist muscles. At this point, the gait cycles were aligned to a grid on the domain $(0, 1)$ to represent the $0^{th}$ to $100^{th}$ percentile of the gait cycle and then interpolated to an evenly spaced grid across subjects.


\subsection{Model formulation}

The \textit{flode} method is a system of differential equations, where equation (\ref{eq:flode_deriv}) represents the model on the scale of the paw velocity, and equation (\ref{eq:flode_mod}) on the scale of the paw position. Because we observe paw position data rather than paw velocities, we estimate parameters using the paw position model. However, we are interested in interpretation on both the position and velocity scales.

In this section we explain our parameter estimation approach. The buffering parameter $\alpha$ will be estimated using nonlinear least squares. Since we observe initial position with error, $Y_i(0)$, we also need to estimate true initial position, $y_i(0)$. The random effects $\delta_i(t)$ and coefficient functions $\mathcal{B}_p(t)$ will be estimated using penalized splines. All parameters will be estimated jointly using the algorithm described in Section \ref{sec:em_algorithm}.

To induce smoothness and reduce dimensionality, the trial-specific random intercepts $\delta_i(t)$ and coefficient functions $\mathcal{B}_p(t)$ are expanded using a fixed B-spline basis, $\bTheta(t)$, of $K$ basis functions  $\theta_1(t),\ldots , \theta_{K}(t)$, such that $\delta_i(t) = \bTheta(t)\bd_i$ and $\mathcal{B}_p(t) = \bTheta(t)\bb_p$, where $\bd_i$, $i \in 1\ldots N$ is a  $K \times 1$ vectors of spline coefficients for the random intercept of the $i$th trial, and $\bb_p$, $p \in 0, \ldots, P$ is a $K \times 1$ vector of spline coefficients for the $p$th coefficient function. Using this representation each forcing function term becomes
\begin{eqnarray*}
\sum_{p=0}^P \int_{s=0}^t e^{-\alpha (t-s)} \cdot x_{ip}(\bs) \cdot \mathcal{B}_p(s) ds &=&\sum_{p=0}^P \int_{s=0}^t e^{-\alpha (t-s)} \cdot x_{ip}(s) \cdot  \Theta(s)\bb_p ds \\[5mm]
&=& \sum_p \left(\int_{s=0}^t \left[\{e^{-\alpha (t-s)}\cdot x_{ip}(s)\}\otimes 1^T_{K}\right] \cdot \Theta(s) ds\right)\bb_p \\[5mm]
&=& \sum_px_{ip}^*(t, \alpha)\bb_p\\
&=& \bx_i^*(t, \alpha)\bb,
\end{eqnarray*}
\noindent
where $\otimes$ denotes the element-wise Kronecker product, and $1_{K}$ is a length $K$ column vector with each entry equal to 1.  We define $\bx_i^*(t, \alpha) = \left\{x_{i0}^*(t, \alpha) | \ldots | x_{iP}^*(t, \alpha) \right\}$ and a $\{K \times (P + 1)\} \times 1$ vector $\bb = \left(\bb_0^T | \ldots | \bb_P^T \right)^T.$ Similarly, the random intercept term becomes

\begin{eqnarray*}
 \int_{s=0}^t e^{-\alpha (t-s)} \cdot \delta_i(s)ds &=&   \int_{s = 0}^t e^{-\alpha (t-s)} \cdot \Theta(s)\bd_i ds\\[5mm]
&=&  \left[\int_{s = 0}^t \left\{ e^{-\alpha (t-s)}\otimes 1^T_{K} \right \}\cdot \Theta(s) ds\right] \bd_i  \\[5mm]
&=& \mathcal{D}^*(t, \alpha)\bd_i,
\end{eqnarray*}
\noindent
Finally, we define $y_{i0}^*(t, \alpha) = y_i(0)e^{-\alpha t}$. An additional aspect of the \textit{flode} model is highlighted in the construction of $\bx_i^*(t, \alpha)\bb$ and  $\mathcal{D}^*(t, \alpha)\bd_i$ above it is necessary to construct design matrices for the spline coefficients via integration, and the result is a nonlinear function of another parameter, $\alpha$. This construction is far from standard in the functional regression literature, and represents a new and unique contribution.

Though the conceptual model is expressed over continuous domain $t$, in practice each trajectory $Y_i$ is observed on the discrete grid, $\bt = \{t_1, t_2, \ldots, t_J\}$, which we assume to be equally spaced and shared across trials. Functions $Y_i(\bt )$ evaluated on this grid are vectors of length $J$, and $\mathcal{D}^*(\bt, \alpha)$ and $x_{ip}^*(\bt, \alpha)$ are $J \times K$ matrices. Then, the function $\bx_i^*(t, \alpha) = \left\{x_{i0}^*(t, \alpha) | \ldots | x_{iP}^*(t, \alpha) \right\}$ is a $J \times \{K \times (P + 1)\}$ matrix. Letting $\bTheta(\bt)$ be a $J \times K$ spline matrix evaluated at $\bt$, then $\delta_i(\bt) = \bTheta(\bt)\bd_i$ and $\mathcal{B}_p(\bt) = \bTheta(\bt)\bb_p$.  We use the notation  $g^*(t, \alpha)$ above to highlight that terms $\bx_i^*(t, \alpha)$,  $\mathcal{D}^*(t, \alpha)$, and $y_{i0}^*(t, \alpha)$ are functions of both time $t$ and the model parameter $\alpha$. However, throughout this section these terms will be used interchangeably with the terms $\bx_i^*$, $\mathcal{D}^*$, and $y_{i0}^*$ for notational simplicity.

We assume both the spline coefficients for the trial-specific intercept, $\bd_i $, and the white noise, $\epsilon_i(t)$, are random and have the distributions $\epsilon_i(t) \sim N(0, \sigma^2I_D)$ and $\bd_i \sim N(0, \Sigma_{K\times K})$, respectively. This induces a conditionally normal distribution on the observed data given the random effects,

$$Y_i| \bd_i \sim N\left(y_{i0}^* + \mathcal{D}^*\bd_i+  \bx_i^*\bb, \sigma^2I_J \right).$$
\noindent
Penalization is a popular technique to avoid overfitting in functional models which we employ here for both random and fixed effect spline coefficients. For fixed effect spline coefficients $\bb_p; p\in 0,\ldots,P$, we assume $\bb_p \sim N(0, \sigma^2_b\mathcal{P}^{-1})$, which introduces a smooth penalty on the coefficient functions. Similarly, we assume the random intercept variance is $\Sigma_{K\times K} = \sigma^2_d\mathcal{P}^{-1}$. Here $\mathcal{P}^{-1}$ is a known penalty matrix that is shared across fixed and random effects to enforce smoothness and regularity.

Using these specifications and evaluating on grid $\bt$ gives the observed data model,
\begin{equation}
\label{eq:observed_mod}
	Y_i(\bt) = y_{i0}^*(\bt, \alpha) +  \mathcal{D}^*(\bt, \alpha)\bd_i +  \bx_i^*(\bt, \alpha)\bb + \epsilon_i(\bt)
\end{equation}
$$\epsilon_i(t) \sim N(0, \sigma^2I_J)$$
$$\bd_i \sim N(0, \sigma^2_d\mathcal{P}^{-1})\mbox{ for }i = 1\ldots N$$
$$\bb_p \sim N(0, \sigma^2_b\mathcal{P}^{-1}) \mbox{ for }p = 1\ldots P.$$

\noindent
We estimate the buffering parameter $\alpha$, variance parameters $\sigma^2$, $\sigma^2_b$, and $\sigma^2_d$, true initial positions $y_i(0)$, and spline coefficients $\bb$ and $\bd_i $ using the expectation-maximization algorithm described in the next subsection. The algorithm incorporates a nonlinear least squares step to optimize the $\alpha$ parameter.


\subsection{EM algorithm for estimating fixed and random effects}
\label{sec:em_algorithm}

We use an expectation-maximization (EM) algorithm to find the maximum likelihood estimates (MLEs) of both fixed and random effects, following precedent from \cite{laird1982} for longitudinal data and \cite{walker1996} for nonlinear mixed models. Our goal is to estimate the experiment-wide fixed effects $\bPhi = \left\{\alpha, \bb, y_i(0), \sigma^2, \sigma^2_d, \sigma^2_b\right\}$ and the random effect spline coefficients $\bd_i$. In the $M$-step of the algorithm we estimate the MLE of the fixed effects when the random effects are observed, $\widehat{\bPhi} = \underset{\bPhi}{\mathrm{argmax}}\{l(\bPhi | Y)\}$, and in the $E$-step we get estimates for the random effects by taking the expectation of the $\bd_i$ under the posterior distribution of $\bd_i$ given the data $Y_i$.

\subsubsection{M-step}

When the random effects $\bd_i$ are known, the MLE of $\bPhi$ maximizes the joint log-likelihood
\begin{eqnarray*}
l(\bPhi) &=& \log p(Y,  \bd ; \bPhi)\\
&=&  \log p(Y|  \bd ; \bPhi) + \log p(\bd; \bPhi) + \sum_{p = 0}^P\log p(\bb_p; \bPhi)\\
&=&  \log p\left\{Y|  \bd ; \alpha, \mathbf{b}, y_i(0), \sigma^2\right\} + \log p(\bd ; \sigma^2_d) + \sum_{p = 0}^P\log p(\bb_p ; \sigma^2_b).
\end{eqnarray*}

\noindent
This leads to the following fixed effects:
\begin{eqnarray*}
\hat{\alpha} &=& \underset{\alpha}{\mathrm{argmin }}<\epsilon^T\epsilon>\\
\widehat{\mathbf{b}} &=& \left\{\bx^{*T}\bx^* + \frac{\sigma^2}{\sigma^2_b} \mathcal{P}  \right\}^{-1}\bx^{*T}\left(Y - y_{0}^* - \bm{\mathcal{D}}^*<\bd> \right)\\
\widehat{y}_i(0) &=& \frac{(e^{-\alpha\bt} )^T \left\{Y_i -  \mathcal{D}^*<\bd_i> - \bx_i^*\bb \right\}}{(e^{-2\alpha\bt})^T 1_{J}}\\
\widehat{\sigma}^2 &=& \frac{<\epsilon^T\epsilon>}{NJ}\\
\widehat{\sigma}^2_d &=& \frac{\sum_i <\bd_i^T\mathcal{P}\bd_i>}{NK}\\
\widehat{\sigma}^2_b &=& \frac{\sum_p \bb_p^T\mathcal{P}\bb_p}{PK}.
\end{eqnarray*}

\noindent
Here $1_{J}$ is a length $J$ column vector with each entry equal to 1. When not indexed by $i$, the vectors $Y$ and $y_0^*$ denote length $NJ$ stacked forms of their trial-specific length $J$ counterparts, $Y_i$ and $y_{i0}^*$. Similarly, $\bd$ is a stacked length $NK$ vector, and $\bx^*$ and $ \bm{\mathcal{D}}^*$ are stacked $NJ \times K$ matrices. The notation $<\ldots>$ represents an expected value. For example, $<\epsilon^T\epsilon>$ denotes the expectation of the residual sum of squares over random effect coefficients $\bd$, and is given by
\begin{eqnarray*}
<\epsilon^T\epsilon> &=& (Y- y_{0}^* - \bm{\mathcal{D}}^*<\bd> -  \bx^*\bb)^T(Y - y_{0}^* - \bm{\mathcal{D}}^*<\bd> -  \bx^*\bb)\\
&=& Y^TY - 2Y^T\left(y_{0}^* + \bm{\mathcal{D}}^*<\bd> + \bx^*\mathbf{b}\right) + y_{0}^{*T}
y_{0}^* + 2 y_{0}^{*T}\left( \bm{\mathcal{D}}^*<\bd> + \bx^*\mathbf{b}  \right) \\
&+& \left( \bx^*\mathbf{b}\right)^T\left( \bx^*\mathbf{b}\right) + 2 \left( \bx^*\mathbf{b}\right)^T \bm{\mathcal{D}}^*<\bd> + <\bd^T\bm{\mathcal{D}}^{*T}\bm{\mathcal{D}}^*\bd>.
\end{eqnarray*}

\noindent
The estimation of $\bd$, $\bd_i^T\mathcal{P}\bd_i$, and $\bd^T\mathcal{D}^{*T}\mathcal{D}^*\bd$ are detailed in the E-step below.

\subsubsection{E-step}

Bayes' rule leads to the posterior distribution of the random intercept coefficients,
$$\bd_i | Y_i \sim N\left(\mathbf{m}_i , \bC\right),$$
\noindent
where
$$\bC = \left\{\frac{1}{\sigma^2_d}\mathcal{P} + \frac{\mathcal{D}^{*T}\mathcal{D}^*}{\sigma^2}\right\}^{-1},$$
\noindent
and
$$\mathbf{m}_i = \frac{\bC \mathcal{D}^{*T}\left(Y_i - y_{i0}^* -  \bx_i^*\bb\right)}{\sigma^2}.$$
\noindent
Then the solutions to $<\bd_i>$, $<\bd_i^T\mathcal{P}\bd_i>$, and $<\bd_i^T\mathcal{D}^{*T}\mathcal{D}^*\bd_i>$ are $\mathbf{m}_i$, $tr(\mathcal{P}\bC) + \mathbf{m}_i^T\mathcal{P}\mathbf{m}_i$, and $tr( \mathcal{D}^{*T}\mathcal{D}^*\bC) + \mathbf{m}_i^T\mathcal{D}^{*T}\mathcal{D}^*\mathbf{m}_i$, respectively, where the notation $tr(A)$ indicates the trace of matrix $A$. We iterate between the $M$ and the $E$-step to obtain a solution. The marginal likelihood of the observed data after integrating out random effects, $l(Y)$,  is used to monitor convergence, which is obtained when the change in $l(Y)$ becomes arbitrarily small.

The random intercept in the \textit{flode} model is included to capture residual within-trial correlation in the paw trajectories. If one assumes that the residuals are uncorrelated, then for each trial $\delta_i(t) = 0$ and the \textit{flode} model simplifies, which allows parameters $\widehat{\bPhi}$ to be maximized directly without the $E$-step.


\subsection{Inference for coefficient functions}

We implement a bootstrap procedure to obtain pointwise confidence intervals for coefficient function estimates $\widehat{\mathcal{B}}_p(t)$ from the \textit{flode} model. To alleviate the computational demand of resampling-based inference, we first fit the \textit{flode} model to the complete data and obtain an estimate $\hat{\alpha}$ from that model fit; this will be treated as a fixed parameter across bootstrap iterations. Specifically, we generate $n_b$ bootstrap samples, fit \textit{flode} to estimate remaining model parameters, and construct coefficient function estimates $\widehat{\mathcal{B}}^{b}_p(t)$ for each bootstrap sample $1 \leq b \leq n_b$. We then calculate pointwise standard error estimates $\hat{se}(\widehat{\mathcal{B}}_p(t))$ by calculating the standard deviation of $\widehat{\mathcal{B}}^{b}_p(t)$ across the bootstrap samples. A Wald-type 95\% confidence interval is then given by $\widehat{\mathcal{B}}_p(t) \pm 1.96 \times \hat{se}(\widehat{\mathcal{B}}_p(t))$. This approach conditions on the estimate of the buffering parameter $\alpha$, which is similar to treating tuning parameters as known in semiparametric inference and works well in our simulations and real data analyses. In cases where $\alpha$ is estimated with notable variance,  a more computationally demanding approach that re-estimates all parameters may be necessary. Inference based on results in semiparametric regression may be possible \citep{rwc2003}, but our preliminary investigations suggested that very large sample sizes would be needed to achieve close to nominal coverage rates.


\subsection{Choice of penalty matrix and initial values}
\label{sec:init}

We choose a penalty matrix commonly used in functional data analysis, $\mathcal{P}$, that enforces smoothness in estimated functions by penalizing the second derivative \citep{eilers1996, goldsmith2016}. To ensure $\mathcal{P}$ is invertible we follow Goldsmith and Kitago and construct $\mathcal{P} = \lambda\mathcal{P}_0 + (1-\lambda)\mathcal{P}_2$, where $\mathcal{P}_0$ and $\mathcal{P}_2$ are the matrices corresponding to the zeroth and second derivative penalties. The $\mathcal{P}_2$ induces smoothness but is not invertible, whereas $\mathcal{P}_0$ is the identity matrix and induces general shrinkage. Combining these two and selecting $0\le \lambda \le 1$ to be small ($\lambda \le 0.01$) ensures that $\mathcal{P}$ is full rank and impose smoothness over shrinkage. A sensitivity analysis comparing \textit{flode} results when $\lambda = 0.001$ and $\lambda = 0.01$ is provided in Figure S.2 of the Web Supplement, and suggests that smaller values of $\lambda$ produce somewhat smoother estimates. Our implementation uses $\lambda = 0.001$.

The EM algorithm detailed in Section~\ref{sec:em_algorithm} converges more quickly when appropriate initial values are provided. To initialize the $\alpha$ parameter, we recommend performing a grid search to find a value $\alpha_0$ that minimizes squared error loss when $\delta_i(t) = 0$, and use this to initialize our full EM algorithm; intuitively, we are choosing an initial buffering parameter after setting random effects to zero and focusing on main effects and the initial position. Our grid search procedure is as follows: a rough grid of $\alpha_0 \in \{0, 20\}$ is defined. For each $\alpha_0$, we obtain an unpenalized ordinary least squares estimate of the fixed effect coefficients, $\widehat{\mathbf{b}}^{OLS}_{\alpha_0} = \left\{\left[\bx^*(t, \alpha_0)\right]^T\bx^*(t, \alpha_0) \right\}^{-1}\left[\bx^*(t, \alpha_0)\right]^T\left[Y - y_{0}^*(t, \alpha_0) \right]$. Then, $\alpha_0$ is chosen to minimize $loss(\alpha_0) = \left[ Y -  y_{0}^*(t, \alpha_0) - \bx^*(t, \alpha_0)\widehat{\mathbf{b}}^{OLS}_{\alpha_0}\right]^T\left[ Y -  y_{0}^*(t, \alpha_0) - \bx^*(t, \alpha_0)\widehat{\mathbf{b}}^{OLS}_{\alpha_0}\right]$. Next, the paw position $y_i(0)$ is initialized using the observed position $Y_i(0)$; random effects $\delta_i(\bt)$ are initialized at 0; and variance parameters $\sigma^2_b$ and $\sigma^2_d$ that control smoothing of the coefficient functions and random effects are initialized at arbitrary high values ($\sigma^2_b = \sigma^2_d = 100$) that correspond to minimal penalization for early EM algorithm iterations. As a sensitivity analyses, we suggest to use random initializations for $\alpha$ drawn from a uniform distribution on $[0,20]$. Results in the supplement explore the number of iterations required for our algorithm to converge and the computation burden in simulated datasets.

\subsection{Implementation}
\label{subsec:implementation}

Our methods are implemented in \texttt{R} and publicly available on \texttt{GitHub}. We use nonlinear least squares to estimate $\alpha$. Specifically, we use the \texttt{optim} function implementation of Brent's method to minimize the squared error loss in Equation (\ref{eq:observed_mod}). On a discrete grid the integrals defined in the $g^*(t, \alpha)$ terms need to be approximated numerically. We use the trapezoidal rule for numeric integration via the \texttt{trapz} function from the \texttt{pracma} package \citep{borchers2021}.


\section{Simulations}
\label{sec:simulations}

We assess the performance of our method using simulations designed to mimic the structure of our motivating data.Data are generated from the \textit{flode} model in Equation (\ref{eq:flode_mod}), varying over the true value of the $\alpha$ parameter to obtain simulation settings that evaluate the sensitivity of our method as $\alpha$ changes. We also generate data from the functional historical regression model in Equation (\ref{eq:fhist_mod}) to study the impact of model misspecification. We compare the results obtained using \textit{flode} to those from functional historical and functional concurrent regression (denoted \textit{fhist} and \textit{fconc} below, respectively).


\subsection{Simulation design and implementation}

Each simulated dataset has $N$ univariate trials $Y_i(\bt), i\in 1\ldots N$ and one forcing function $\bx_1(t)$. All trials share the same equally-spaced grid, $\bt \in [0, 1]$, of length $J = 50$. Initial positions $y_i(0)$ are sampled from $N(0, 5)$ for each trial. Forcing functions take the form $\bx_{i1}(\bt) = scale_i \times \sin(\pi \bt + shift_i)$, where  $scale_i$ and $shift_i$ are randomly-drawn, trial-specific scale and shift parameters. Random intercepts $\delta_i(\bt) = \bTheta(\bt)\bd_i$ are constructed using a cubic B-spline basis $\bTheta(\bt)$ with 10 degrees of freedom and spline coefficients $\bd_i$, which are drawn from  $\bd_i \sim N(0, \sigma^2_d I_{10})$ with $\sigma^2_d = 50$. Measurement errors $\epsilon_i(\bt)$ are drawn from  $\epsilon_i(\bt) \sim N(0, \sigma^2 I_{J})$, where $\sigma^2 = 0.1$, an amount of residual variance which is comparable to that seen in our motivating data.

For data generated under the \textit{flode} model, we construct the population intercept $\mathcal{B}_0(t)$ and coefficient function $\mathcal{B}_1(t)$ to be smooth functions over $t \in [0,1]$ and choose the buffering parameter $\alpha \in (0.1, 0.5, 1, 2, 4, 8, 12)$. For data generated under the \textit{fhist} model, we construct a coefficient surface that exhibits a strong, local effect centered around $t = 0.25$ and $s = 0.75$; this surface was chosen to be very distinct from a surface that can be well-estimated using \textit{flode}. A visual comparison of the \textit{fhist} coefficient surface and the surfaces induced by the \textit{flode} specification is available in Figure \ref{fig:sim_surfaceErr}. Under each of these data generating mechanisms, we set $N= 100$ and simulate 50 datasets. We additionally simulate 30 datasets for each combination of $N \in (10, 50, 100, 500)$ and $\alpha \in (0.5, 4, 12)$ to understand our method's performance across different numbers of trials. Lastly, we generate 500 datasets with $\alpha = 4$ and $N= 100$, obtain bootstrap 95\% confidence intervals for each dataset, and evaluate mean pointwise coverage of these intervals.

The \textit{flode} model is implemented as described in Section~\ref{sec:methods}, using the implementation in Section~\ref{subsec:implementation}. We use $K = 20$ B-spline basis functions and initialize $\alpha$, $y_i(0)$, and random effects $\delta_i(\bt)$ following the recommendations in Section~\ref{sec:init}. The functional historical model is implemented using the \texttt{pffr()} function from the \texttt{refund} package in \texttt{R} \citep{refund}. Our first implementation of this approach suggested that the default number of basis functions was insufficient to capture the complexity in our coefficient surfaces, and was increased to a $15 \times 15$ tensor product basis. Although it is not used as a data generating mechanism, some comparisons to the functional concurrent regression are informative. We implemented functional concurrent regression using the \texttt{vbvs.concurrent} package in \texttt{R}, with default values for all function arguments \citep{goldsmith2016vbvs}.


\subsection{Evaluation metrics}
\label{sec:metrics}

We evaluate our approach in terms of accuracy for estimating \textit{flode} parameters, particularly the buffering parameter $\alpha$ and the coefficient functions $\mathcal{B}_p(t)$, when that model is the true data generating mechanism. The differential form of the \textit{flode} model in Equation (\ref{eq:flode_deriv}) is, fundamentally, the model we propose; this also provides the clearest mechanism for interpreting coefficients through the forcing function's influence on paw velocity. Competing methods do not estimate these parameters, and are therefore not evaluated for estimation accuracy. Below we quantify recovery of the true values of $\alpha$, $\mathcal{B}_0(t)$, and $\mathcal{B}_1(t)$ across simulation scenarios. For $\alpha$ estimation accuracy is quantified by $error = \alpha^{true}-\hat{\alpha}$, and for each coefficient function we calculate integrated error defined by $IE = \int_t \{\mathcal{B}_p^{true}(t) - \hat{\mathcal{B}}_p(t) \}dt$. Note that we calculate integrated errors rather than integrated squared errors to assess bias (i.e. consistent over or under estimation) across the range of true $\alpha$ values. Performance of confidence intervals is evaluated through average integrated pointwise coverage of $95\%$ confidence intervals, defined as
$\frac{1}{500}\sum_{l = 1}^{500}\int_{0}^{1} I\left\{\mathcal{B}^{true}_1(t) \in \widehat{\mathcal{B}}^{l}_1(t) \pm 1.96 \times \hat{se}(\widehat{\mathcal{B}}^{l}_1(t))\right\}\, dt$, where $\widehat{\mathcal{B}}^{l}_1(t)$ is the estimate of $\mathcal{B}_1(t)$ for the $l^{th}$ bootstrapped sample and $I\{\}$ is an indicator function taking the value $1$ when the evaluated expression is true and $0$ otherwise.

The integrated form of the \textit{flode} model in Equation (\ref{eq:flode_mod}) yields a coefficient surface as a byproduct of $\alpha$ and the coefficient functions $\mathcal{B}_p(t)$. While not a quantity of direct interest, this surface is useful for comparison with the functional historical regression in Equation (\ref{eq:fhist_mod}). We denote the ``true'' surface as $\beta_1^{true}(s, t) = e^{-\alpha^{true}(t-s)} \mathcal{B}_1^{true}(s) \mathcal{I}(s < t)$ and compare it with the estimated surfaces from \textit{flode} and \textit{fhist}. Let the estimated surface from \textit{flode} be $\hat{\beta}_1^{flode}(s, t) = e^{-\hat{\alpha}(t-s)} \hat{\mathcal{B}}_1(s)\mathcal{I}(s < t)$ and the estimated \textit{fhist} surface $\hat{\beta}_1^{fhist}(s, t)$. Surface recovery accuracy is quantified using the integrated squared error (ISE), where for \textit{flode} $ISE = \int_t \int_s \left\{ \beta_1^{true}(s, t)  - \widehat{\beta}_1^{flode}(s, t) \right\}^2 ds dt$ and for \textit{fhist} $ISE = \int_{0}^{1} \int_s \left\{\beta_1^{true}(s, t)  - \widehat{\beta}_1^{fhist}(s, t) \right\}^2 ds dt$.

Lastly, we compare \textit{flode}, \textit{fhist} and \textit{fconc} in terms of prediction accuracy. For each true data generating mechanism, we produce an evaluation dataset of 10,000 observations containing forcing functions $\bx_{i1}(\bt)$ and position trajectories $Y_i(\bt)$. Model parameters are estimated from a training dataset, and we then we obtain mean absolute prediction errors defined as $MAPE = \frac{1}{10000} \sum_{i=1}^{10000}\int_t |\hat{Y}_i^{predict}(t)-Y_i(t)|dt$, where $\hat{Y}_i^{predict}$ is the model-based prediction of the response and the sum is taken over the evaluation dataset; this is repeated for \textit{flode} to \textit{fhist} and \textit{fconc} for each of the 50 training datasets. The distribution of MAPE values is used to understand whether accurate position trajectories can be produced when the model is mispecified.


\subsection{Simulation results}

\begin{figure}[h]
  \centering
     \begin{tabular}{cc}
      	\includegraphics[width= .9\textwidth]{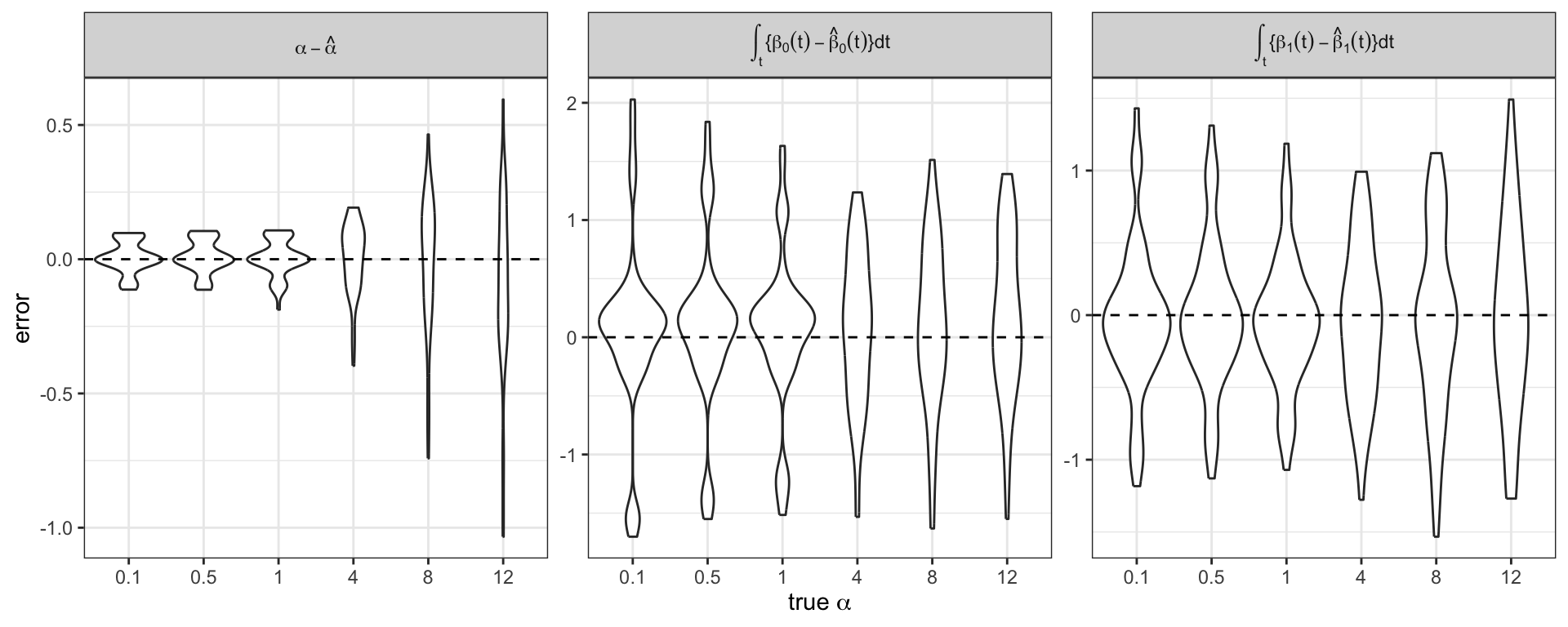}
     \end{tabular}
     \caption{Distribution of estimation error in $\widehat{\alpha}$ (left panel) and integrated error in $\widehat{\mathcal{B}}_0(t)$ and $\widehat{\mathcal{B}}_1(t)$ (middle and right panels) from the \textit{flode} model across 50 simulated datasets with true $\alpha \in (0.1, 0.5, 1, 2, 4, 8, 12)$ and $N = 100$.}
    \label{fig:sim_alpha}
\end{figure}

Figure \ref{fig:sim_alpha} quantifies estimation error for \textit{flode} parameters ${\alpha}$, ${\mathcal{B}}_0(t)$, and $\widehat{\mathcal{B}}_1(t)$ in the left, middle, and right panels, respectively. Distributions of error and integrated error (IE) values across 50 simulated datasets with true values of $\alpha \in (0.1, 0.5, 1, 2, 4, 8, 12)$ are displayed. Our \textit{flode} implementation accurately recovers the true value of $\alpha$ across varying $\alpha$ values, with little or no evidence of bias in the estimation procedure. The variability of \textit{flode} estimate $\widehat{\alpha}$ increases as the true value of $\alpha$ increases; this is perhaps not surprising since $\alpha$ is non-negative and the range of possible values is narrower when $\alpha$ is small. For coefficient functions $\mathcal{B}_0(t)$ and $\mathcal{B}_1(t)$, the IE distributions are relatively stable across true $\alpha$ values. These distributions are also centered around zero, suggesting little estimation bias. In the simulations that examined coverage of confidence intervals, we found that $95\%$ intervals for ${\mathcal{B}}_0(t)$ and $\widehat{\mathcal{B}}_1(t)$ yielded average integrated pointwise coverages across datasets of 0.933 and 0.934, respectively, which indicates our method for inference has satisfactory numerical properties.

\begin{figure}[h]
  \centering
     \begin{tabular}{cc}
      	\includegraphics[width= .95\textwidth]{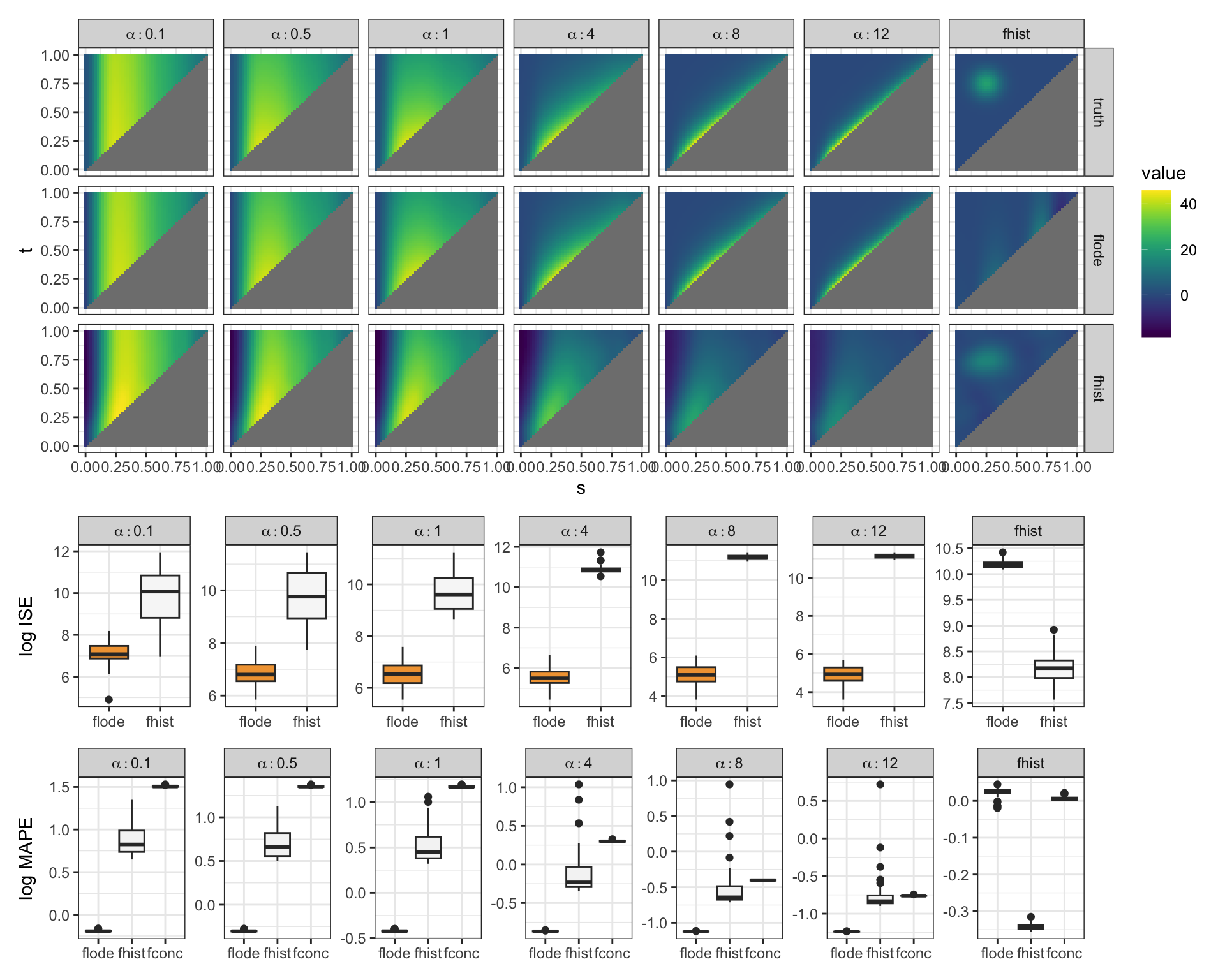}
     \end{tabular}
     \vspace*{5mm}
     \caption{Performance of \textit{flode}, \textit{fhist}, and \textit{fconc} estimates in simulated datasets generated from both \textit{flode} and \textit{fhist} data generating models. Top three rows compare surface $\beta(t,s)$ estimates from \textit{flode} and \textit{fhist} across datasets simulated from a \textit{flode} model using values of $\alpha \in (0.1, 0.5, 1,2,4,6,8,10,12)$ (left 6 columns) and a dataset generated from an \textit{fhist} model with a true surface that cannot be well-estimated by \textit{flode}. \textit{flode} performs well when data is generated from a \textit{flode} model, and when data is generated from an \textit{fhist} model \textit{fhist} outperforms \textit{flode}. Fourth row shows integrated squared surfaces errors defined in Section \ref{sec:metrics} from \textit{flode} and \textit{fhist}. Bottom row shows prediction accuracy of \textit{flode}, \textit{fhist}, and \textit{fconc} models as measured by log mean absolute prediction errors ($\log$MAPE) from 50 simulated training datasets. Prediction errors for each panel were calculated on a single simulated test dataset with 10,000 functional observations.}
    \label{fig:sim_surfaceErr}
\end{figure}

Figure \ref{fig:sim_surfaceErr} contains comparisons \textit{flode}, \textit{fhist}, and \textit{fconc} across datasets generated under the \textit{flode} model with values of $\alpha \in (0.1, 0.5, 1,2,4,6,8,10,12)$, as well as datasets generated under the \textit{fhist} model. The top rows of Figure~\ref{fig:sim_surfaceErr} show the true coefficient surfaces for each data generating mechanism (denoted ``truth") as well as examples of surfaces obtained by \textit{flode}, \textit{fhist} for a single dataset with $N = 100$. The true coefficient surfaces further illustrate the effect of the buffering parameter $\alpha$. When $\alpha$ is small, the coefficient surface at a fixed value of $s$ is nearly constant over $t$ and the impact of a forcing function persists through time, but when $\alpha$ is large the coefficient surface is almost zero when $t > s$. For datasets generated under the \textit{flode} model, the estimated surfaces obtained through \textit{flode} are nearly identical to the true surface. Surfaces obtained by \textit{fhist} capture the correct general shape, but the more flexible model structure results in visually poorer estimation. As expected, for the dataset generated under the \textit{fhist} model, the \textit{flode} model is unable to capture the main features while \textit{fhist} performs well.

The middle panels of Figure~\ref{fig:sim_surfaceErr} show the distribution of $\log ISE$ values for surface estimates obtained through \textit{flode} and \textit{fhist} across all simulated datasets. The results here are consistent with observations regarding surface estimates for single datasets. Specifically, across values of $\alpha$, \textit{flode} outperforms \textit{fhist} in terms of the $\log ISE$. At low values of $\alpha$, the difference in performance between the methods is smaller, although $\log ISE$ variability for \textit{fhist} is high when $\alpha < 1$. Values of $\log ISE$ for \textit{flode} decrease as $\alpha$ increases. Meanwhile, when \textit{fhist} is the true data generating model, estimates obtained through \textit{fhist} substantially outperform estimates from \textit{flode}. For each data generating mechanism, the distribution of $\log$MAPE values across simulated datasets for \textit{flode}, \textit{fhist}, and \textit{fconc} appear in the bottom panels of Figure~\ref{fig:sim_surfaceErr}. When data are generated under the \textit{flode} model, the improved accuracy of coefficient surface estimates obtained through \textit{flode} compared to \textit{fhist} translates to lower $\log$MAPE values. Results from \textit{fconc} are generally worse than either \textit{flode} or \textit{fhist}, although this gap narrows as $\alpha$ increases and the model more closely resembles a concurrent regression. As expected, when data are generated under the \textit{fhist} model both \textit{flode} and \textit{fconc} perform notably worse than \textit{fhist}.

Additional simulation results are reported in the Web Supplement. Figure S.1 explores convergence rates of the EM algorithm described in Section~\ref{sec:em_algorithm} and computation times for both \textit{flode} and \textit{fhist} across values of $\alpha$. These results indicate that \textit{flode} converges fastest for higher $\alpha$ values, and that for the settings we consider \textit{flode}'s computation time is similar to or faster than that of \textit{fhist}. Figure S.2 examines sensitivity to the parameter $a$ used to construct the penalty matrix $\mathcal{P}$; choosing $a =0.001$ induces more second derivative penalization and results in somewhat smoother estimates compared to $a = 0.01$, and that setting $a =0.001$ reduces sensitivity to the choice of the basis dimension $K$. Figure S.3 shows distributions of estimation error in $\widehat{\alpha}$ (left column) and integrated error in $\widehat{\mathcal{B}}_0(t)$ and $\widehat{\mathcal{B}}_1(t)$ (middle and right columns) from the \textit{flode} model across 50 simulated datasets each with true $\alpha \in (0.5, 4, 12)$ and $N \in (10, 50, 100, 200)$. Across simulation scenarios, \textit{flode} parameter estimates appear to be relatively unbiased. Unsurprisingly, and number of trials $N$ increases, variance of parameter estimates decreases.


\section{Data Analysis}
\label{sec:results}

We now turn our attention to the analysis of the mouse gait data introduced in Section \ref{sec:data}. This dataset consists of paw upward position and muscle activation signals recorded simultaneously during each of 161 gait cycles as a single mouse trotted on a treadmill. Because the mouse paused or engaged in other behaviors, gait cycles and not necessarily consecutive. Position and muscle activation were obtained and processed using standard techniques, and functional data were interpolated to a shared grid of 100 measurements per gait cycle. Each gait cycle is defined to begin when the paw has come to rest at the conclusion of the previous cycle, so that muscle activation preceding paw lift is observed in the same gait cycle. Our goal is to model the impact of muscle activation in the biceps and triceps on paw position in order to better understand neuromuscular control during locomotion.

\begin{figure}[h]
  \centering
     \begin{tabular}{cc}
      	\includegraphics[width= .95\textwidth]{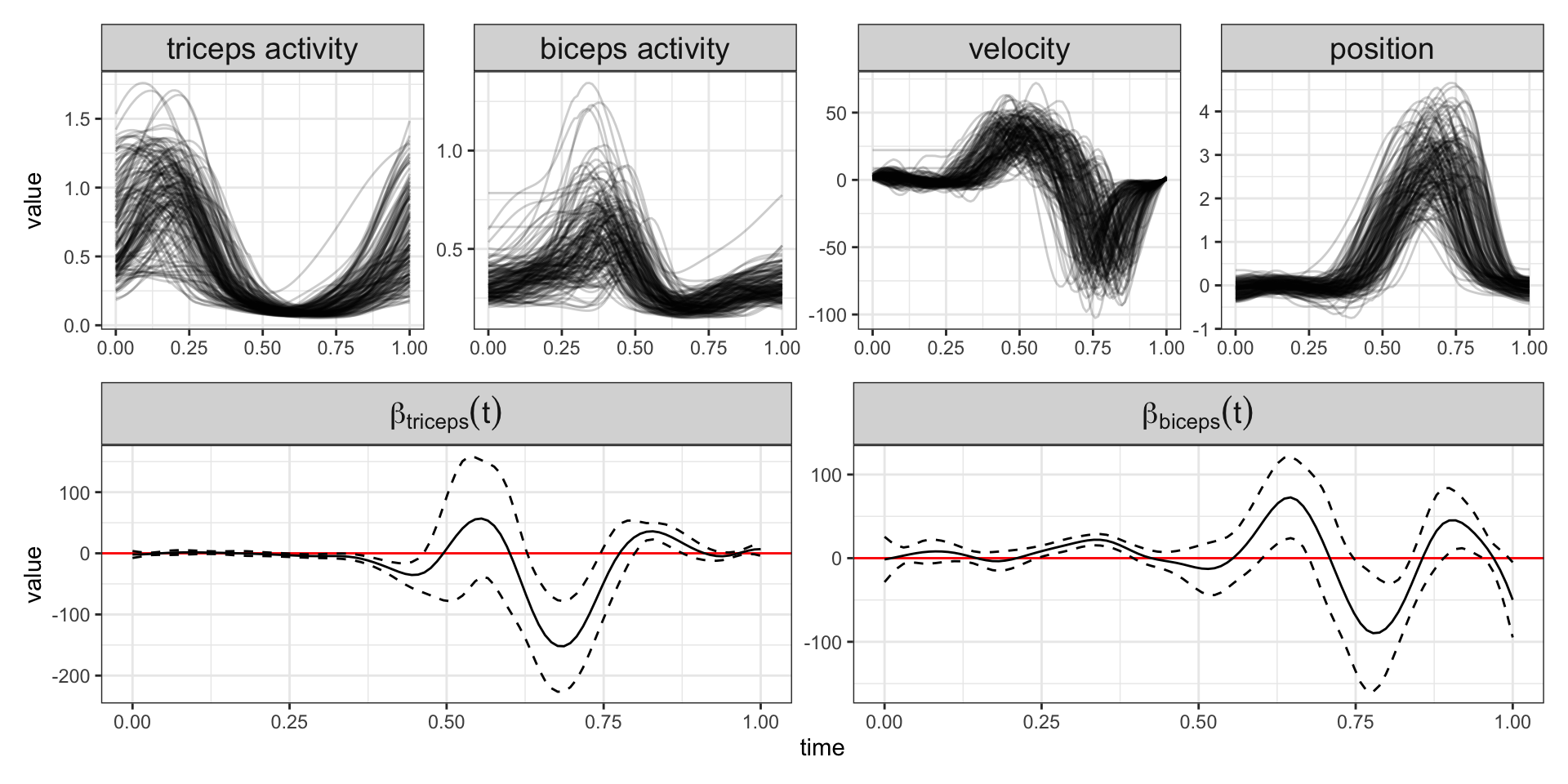}
     \end{tabular}
     \vspace*{5mm}
     \caption{First row: smoothed triceps activation, smoothed biceps activation, paw upward velocity, and paw position for all 161 gait cycles.
     Second row: Fitted coefficient functions and bootstrapped 95\% confidence intervals for the biceps and triceps forcing functions.
     }
    \label{fig:paw_betas}
\end{figure}

Data obtained from five gait cycles in displayed in Figure~\ref{fig:flode_data}. In addition to muscle activation and paw position this Figure shows numerically-obtained velocity curves; these are used to illustrate the biomechanical interpretation of the proposed \textit{flode} model, although the estimation procedure does not use these directly. Data from all gait cycles are shown in the top row of Figure~\ref{fig:paw_betas}.

In this Section we begin by applying the \textit{flode} model as described in Section~\ref{sec:methods} and interpreting the estimated parameters in the context of the motivating experiment. We then compare the results obtained from \textit{flode} to those from functional historical regression (\textit{fhist}) and functional linear concurrent regression (\textit{fconc}) in terms of predictive performance. To this end, we performed ten fold cross-validation and summarize predictive performance using mean absolute prediction error (MAPE). For each \textit{flode} and \textit{fconc}, we set the number of spline bases used to model coefficient functions to $K = 20$ and for \textit{fhist} we use a $15 \times 15$ tensor product basis to model the coefficient surface. Initial values for \textit{flode} are set using the guidance in Section~\ref{sec:init}. Other than the number of basis functions, default arguments are used for the \texttt{pffr()} and \texttt{vbvs.concurrent} implementations of \textit{fhist} and \textit{fconc}.


\subsection{Interpretation of \textit{flode} parameters}

Panels in the bottom row of Figure \ref{fig:paw_betas} show estimated coefficient functions $\hat{\mathcal{B}}_p(t)$ for the triceps and biceps from the fitted \textit{flode} model, along with their 95\% confidence intervals. Recalling the differential form of the \textit{flode} model in Equation~(\ref{eq:flode_deriv}), these coefficients characterize the instantaneous effect of the forcing functions on the paw velocity which, in turn, affects paw position. For the triceps, there are significant negative associations between activation and velocity for $t \in [0.24, 0.46]$ and $t \in [0.64, 0.74]$, suggesting that higher triceps activation is associated with lower velocities. The estimated effect size is smaller in magnitude in the first interval that in the second, although variability in triceps activation is higher. The first interval coincides with increases in velocity and subsequent increases in position; higher than average triceps activation indicates a delay in the process leading to paw lift (i.e. low velocity for longer during the gait cycle). The second interval, meanwhile, coincides with decreases in velocity as the paw approaches and lands on the treadmill; higher triceps activation in this interval reflects more rapid deceleration as the paw approaches and lands on the treadmill.

Biceps activation has a similar biomechanical interpretation. For example, there are significant positive associations between biceps activation and paw velocity for $t \in [0.25, 0.39]$ and $t \in [0.61, 0.67]$, as well as a negative association for $t \in [0.75, 0.83]$. In the first interval, which coincides with the onset of the process leading to paw lift, increased biceps activation corresponds to increased positive velocity. Similarly, in the second interval, higher than average biceps activation is associated with increased paw velocity, which may reflect gait cycles in which the paw remains lifted for longer periods. More generally, we note that the coefficients for the biceps and triceps are significant in broadly similar regions but have opposite signs. This potentially reflects the opposing biomechanical outputs of flexor and extensor muscles on paw velocity and position.

The buffering parameter, $\alpha$, has an estimated value of $\widehat{\alpha} = 3.22$. This imposes a relatively high degree of buffering, suggests that the effects of triceps and biceps activation on paw velocity impacts paw position in a persistent way.  Put differently, the \textit{flode} model indicates that paw position depends on the current and recent paw velocity, and that velocity is directly affected by muscle activation.


\subsection{Comparison of \textit{flode} and standard functional regression models}

As detailed in Section~\ref{sec:fda}, functional historical (\textit{fhist}) and functional linear concurrent regression (\textit{fconc}) are alternative frameworks that could be used to model the impact of muscle activation on paw position. In contrast to \textit{flode}, these models do not posit that the impact of muscle activation on position takes place through changes in velocity. Instead, they model paw position as the direct outcome of muscle activation, and can serve as plausible alternatives to \textit{flode}.

We compare \textit{flode} to \textit{fhist} and \textit{fconc} in terms of predictive accuracy, using 10-fold cross validation to obtain out-of-sample mean absolute prediction errors (MAPEs). Using this evaluation metric, \textit{flode} had best overall predictive performance ($MAPE = 0.293$), followed by \textit{fconc} ($MAPE = 0.311$) and then \textit{fhist} ($MAPE = 0.398$). The comparison between \textit{flode} and \textit{fhist} lends credibility to the \textit{flode} specification and interpretation: as shown in simulations, a setting in which \textit{flode} is obviously misspecified can result in poor prediction accuracy. Meanwhile, the relatively poor performance of \textit{fhist} may reflect the increased flexibility of this model and a tendency to overfit in this setting. This is supported by visual inspections of the \textit{flode} and \textit{fhist} coefficient surfaces (Figure S.4); these have qualitatively similar shapes but the \textit{fhist} surface is less smooth. The performance of \textit{fconc} suggests that this model can produce reasonable fitted values in this setting, and may reflect the relatively straightforward model parameterization and estimation. At the same time, the interpretation of \textit{fconc} -- that muscle activation has a concurrent effect on position -- is less biomechanically plausible than the interpretation of \textit{flode} discussed above. We also note that the value of the buffering parameter $\alpha$ in the \textit{flode} fit suggests a historical effect.




\section{Discussion}
\label{sec:discussion}

We present \textit{flode}, a novel nonlinear regression model that has context in both functional data analysis and systems of ordinary differential equations. Drawing from both of these literatures is necessitated by our application; the differential equations formulation of our model allows for an interpretation of our gait data as trajectories whose speed and position are dynamically influenced by inputs from the biceps and triceps muscles, and tools from functional data analysis allow us to efficiently model observations that are trajectories while incorporating smoothness in the coefficient functions. Though we are motivated by a specific biokinematics application, our methods are broadly useful for dynamical systems of inputs and outputs where the outputs are functions over time. Our method compares favorably with historical functional regression and functional linear concurrent regression in the real data and simulation settings we examined, and produces interpretable results for our motivating data. Our methods, as well as \texttt{R Markdown} documents that detail our analysis pipeline and fully reproduce simulation Figures and results, are publicly available in \texttt{R} on \texttt{GitHub}.

There are several important directions for future work in this area. A study on the asymptotics of the coefficients estimated in this model so that large sample confidence intervals and hypothesis tests can be computed would help researchers draw inferences about the relationships between inputs and outputs of the dynamical system. Extensions to include more complex systems of ordinary differential equations, including higher order and non-linear ODEs would increase the flexibility of our modeling framework and allow for the study of a larger class of repeated measurements of dynamical systems.

We developed the \textit{flode} model to better reflect our current understanding of biological and biomechanical processes for our application in which muscle activation impacts paw position through changes to velocity.
Our model provides an interpretable description of this mechanism. In addition to the methodological extensions above, including to higher-order differential equations and non-linear models, we are interested in extensions that will allow the interrogation of more complex systems. For example, there is evidence that neural firing in the motor cortex controls muscle output which in turn affects the velocity and position of the paw during a gait cycle. Each step in this process can be viewed as a component in a system that reacts to ongoing inputs, and developing a statistical framework that leverages both functional data analysis and differential equations would contribute to the ultimate goal of understanding how neural signals drive or modify locomotion under changing circumstances.

%
%

\begin{acks}[Acknowledgments]
The authors would like to thank the anonymous referees, an Associate
Editor and the Editor for their constructive comments that improved the
quality of this paper.
\end{acks}

\begin{funding}
The first author was supported by NSF Grant DMS-??-??????.

The second author was supported in part by NIH Grant ???????????.
\end{funding}




\bibliographystyle{imsart-nameyear} 
\bibliography{biblio}       


\end{document}